\documentclass[twocolumn,aps,pra,showpacs]{revtex4}

\usepackage{graphicx}
\usepackage{dcolumn}
\usepackage{bm}
\usepackage[latin1]{inputenc}

\def\be{\begin{equation}}
\def\ee{\end{equation}}
\def\bea{\begin{eqnarray}}
\def\eea{\end{eqnarray}}

\begin{document}

\title{Bell's inequality for $n$ spin-$s$ particles}
\author{Ad\'{a}n Cabello}
\email{adan@us.es}
\affiliation{Departamento de F\'{\i}sica Aplicada II,
Universidad de Sevilla,
41012 Sevilla, Spain}
\date{\today}


\begin{abstract}
The Mermin-Klyshko inequality for $n$ spin-${1 \over 2}$ particles and two
dichotomic observables is
generalized to $n$ spin-$s$ particles and two maximal observables.
It is shown that some multiparty multilevel
Greenberger-Horne-Zeilinger states [A. Cabello, Phys. Rev. A {\bf
63}, 022104 (2001)] maximally violate this inequality for any $s$.
For a fixed $n$, the magnitude of the violation is constant for any
$s$, which provides a simple demonstration and generalizes the conclusion
reached by Gisin and Peres for two spin-$s$ particles in the
singlet state [Phys. Lett. A {\bf 162}, 15 (1992)]. For a fixed
$s$, the violation grows exponentially with $n$, which provides a
generalization to any $s$ of Mermin's conclusion for $n$ spin-${1
\over 2}$ particles [Phys. Rev. Lett. {\bf 65}, 1838 (1990)].
\end{abstract}

\pacs{03.65.Ud,
03.65.Ta}
\maketitle


\section{Introduction}


Einstein, Podolsky, and Rosen (EPR) \cite{EPR35} believed that
the results of experiments on
a local subsystem of a composite physical
system which can be predicted with certainty from the results of
local experiments in other regions would be determined by the local
properties of the subsystem. However, the violation of Bell's
inequality by quantum mechanics \cite{Bell64} meant a spectacular
departure from EPR's point of view. According to quantum mechanics, the
results of local experiments cannot be described in terms of
classical local properties.

On the other hand, it was commonly accepted that classical
properties would emerge for large quantum systems. The adjective
``large'' usually means either systems composed of many particles or
systems with a high number of internal
degrees of freedom.
Early violations of Bell's inequalities
\cite{Bell64,CHSH69} involved pairs of spin-${1 \over 2}$
particles in the singlet state \cite{Bohm51}. However, the EPR
argument is also applicable to pairs of spin-$s$ particles in the
singlet state or to systems of $n$ spin-${1 \over 2}$ particles in
Greenberger, Horne, and Zeilinger (GHZ) states \cite{GHZ89}.
Violations of Bell's inequalities for the two spin-$s$ singlet state
have been extensively discussed
\cite{Mermin80,MS82,GM82,GM83,Ogren83,BC88,SS90,Ardehali91,Peres92,GP92}
and have stimulated some recent experiments for $s=1$
\cite{LHB01,HLB02}.
On the other hand, violations of Bell's
inequalities for $n$ spin-${1 \over 2}$ particles have attracted
much attention
\cite{Mermin90b,RS91,CRB91,Ardehali92,BK93,BM93,ZK97,GB98,WW00,WW01}. However, a
study of Bell's inequalities for systems of $n$ spin-$s$
particles and the limit of both $n \rightarrow \infty$ and $s
\rightarrow \infty$ was still missing.

In order to place our discussion in a suitable context,
we shall review
some of the earlier violations of Bell's inequalities
for two spin-$s$ particles and for $n$ spin-${1 \over 2}$
particles.

Mermin \cite{Mermin80} showed that a pair of spin-$s$ particles
in the singlet state violates a particular Bell's inequality involving
four local spin component observables
${\vec S}_1 \cdot {\hat a}$, ${\vec S}_1 \cdot {\hat b}$,
${\vec S}_2 \cdot {\hat b}$, and ${\vec S}_2 \cdot {\hat c}$.
He found that the range of settings for which
the violation occurs
vanishes as $1/s$ when $s \rightarrow \infty$.
Subsequently, however, Mermin and Schwarz \cite{MS82} found evidence that
this vanishing might be peculiar to the
chosen inequality (see also \cite{SS90,Ardehali91}).

\"{O}gren \cite{Ogren83} studied the original Bell's inequality \cite{Bell64}
for three different ways of defining dichotomic observables from
${\vec S}_1 \cdot {\hat a}$, ${\vec S}_1 \cdot {\hat b}$, ${\vec
S}_2 \cdot {\hat b}$, and ${\vec S}_2 \cdot {\hat c}$. He found
that the range of settings for which the singlet state of two
spin-$s$ particles violates Bell's inequality is of the same
magnitude, at least for small $s$, and larger than those obtained
in Ref.~\cite{Mermin80}.

Peres \cite{Peres92} and Gisin and Peres \cite{GP92}
found dichotomic operators
such that two spin-$s$ particles in the singlet state violate
the Clauser-Horne-Shimony-Holt \cite{CHSH69} (CHSH) inequality and that
the magnitude of the violation
(that is, the ratio of the quantum
correlation to the maximal classical one) tends to a constant
\cite{Peres92} or is constant \cite{GP92} for any $s$.

An experimental violation of Bell's inequalities
for an optical analog of the singlet state of two spin-1 particles
has been recently reported in Ref.~\cite{HLB02}.

On the other hand, Mermin \cite{Mermin90b} has shown that
the correlations found by $n$ spacelike separated observers
who share $n$ spin-${1 \over 2}$ particles in a GHZ state
maximally violate a Bell's inequality involving two local
spin component observables per particle by
a factor that increases exponentially with $n$.
Mermin's inequality for $n$ spin-${1 \over 2}$ particles
distinguishes between the $n$ even and odd cases.
Ardehali \cite{Ardehali92} derived a similar inequality that
leads to a higher violation for even $n$.
Finally, Belinsky and Klyshko \cite{BK93} proposed an
elegant single inequality that leads to a maximal violation for arbitrary $n$.
This inequality is mostly referred to as the Mermin-Klyshko inequality.

The structure of this paper is as follows:
In Sec.~II we introduce a generalization for any spin of
the Mermin-Klyshko inequality
using two maximal observables (i.e., represented by
nondegenerated operators) per particle.
In Sec.~III we show that maximally entangled states of two
spin-$s$ particles and some multiparticle multilevel GHZ states
defined in Ref.~\cite{Cabello01} maximally violate the inequality presented in Sec.~II.

In Sec.~IV we present the conclusions of our research:
On one hand, we reach Gisin and Peres's conclusion
in Ref.~\cite{GP92}, namely that
for two particles in a maximally entangled state the ratio of the quantum
correlation to the maximal classical one is constant as $s$ grows.
Moreover, we
extend Gisin and Peres's conclusion to systems of three or more particles.
On the other hand, we
generalize to any $s$ Mermin's conclusion in Ref.~\cite{Mermin90b} that the ratio of the
quantum correlation to the maximal classical one grows exponentially with the
number of particles.
In addition, the inequality presented in Sec.~II would allow us to
translate the proofs of
Bell's theorem without inequalities for multiparticle multilevel GHZ
states introduced in Ref.~\cite{Cabello01}
into feasible experimental tests.


\section{The Mermin-Klyshko inequality for {\em n} spin-{\em s} particles}


Let us consider a system with $n \ge 2$ distant spin-$s$ particles, 1, \ldots, $n$
shared by $n$ distant observers which perform spacelike local experiments,
chosen between $A_j^{(s)}$ and $B_j^{(s)}$, on his/her particle $j$.
Let us choose units in which $\hbar =1$ and
let $A_j^{(s)}$ and $B_j^{(s)}$ be physical observables
on particle $j$ taking values $-s$, $-s+1$, \ldots, or $s$.

The correlation $A_1^{(s)}
\ldots A_n^{(s)}$ of $A_1^{(s)}$, \ldots, $A_n^{(s)}$ is defined as
\bea
A_1^{(s)} \ldots A_n^{(s)} & = & \nonumber \\
& &
\!\!\!\!\!\!\!\!\!\!\!\!\!\!\!\!\!\!\!\!\!\!\!\!\!\!\!\!\!\!\!\!\!\!\!\!\!\!
\sum_{m_1,\ldots,m_n=-s}^{s}
\!\!\!\!\!\!\!\!\!m_1 \ldots m_n P(A_1^{(s)}\!\!=m_1,\ldots,A_n^{(s)}\!\!=m_n),
\eea
where $P(A_1^{(s)}=m_1,\ldots,A_n^{(s)}=m_n)$
is the joint probability of obtaining
$A_1^{(s)}=m_1$, \ldots, and $A_n^{(s)}=m_n$
when $A_1^{(s)}$, \ldots, and $A_n^{(s)}$ are measured.

Let us consider the linear combination of $2^{2 f\left({n / 2}\right)}$
correlations, where $f(x)$ is the greatest integer less than or equal to $x$,
defined recursively by
\be
M_n^{(s)}=M_{n-1}^{(s)}
\left(A_n^{(s)}+B_n^{(s)}\right) +
K_{n-1}^{(s)}
\left(A_n^{(s)}-B_n^{(s)}\right),
\ee
letting $M_1^{(s)}=A_1^{(s)}$, and $K_n^{(s)}$ being
the same as $M_n^{(s)}$ but exchanging
the $A$'s for $B$'s.

In particular,
\be
M_2^{(s)} =
A_1^{(s)} A_2^{(s)}+A_1^{(s)} B_2^{(s)}
+ B_1^{(s)} A_2^{(s)}-B_1^{(s)} B_2^{(s)}
\ee
and
\bea
M_3^{(s)} & = & 2 \left(
A_1^{(s)} B_2^{(s)} B_3^{(s)} +
B_1^{(s)} A_2^{(s)} B_3^{(s)} \right. \nonumber \\
& & \left. +
B_1^{(s)} B_2^{(s)} A_3^{(s)} -
A_1^{(s)} A_2^{(s)} A_3^{(s)} \right).
\eea

In any theory in which local variables of particle $j$ determine
the results of local observables $A_j^{(s)}$ and $B_j^{(s)}$, the absolute value
of $M_n^{(s)}$ is bound as follows:
\be
\left|M_n^{(s)}\right| \le 2^{n-1} s^n.
\label{Mermininequalitys}
\label{Mermins}
\ee
This is the generalization to spin-$s$ of the Mermin-Klyshko inequality.
If $A_j$ and $B_j$ are observables taking values $-1$ or $1$ (i.e., for $s=1$),
or for $s={1 \over 2}$ and choosing units in which $2 \hbar =1$, then
we obtain the Mermin-Klyshko inequality \cite{BK93}.
If, in addition, $n$ is odd and greater than 3, then
(up to a factor $2^{f\left({(n-1) / 2}\right)})$ we obtain
Mermin's inequality \cite{Mermin90b}.
If $n=2$ we obtain the CHSH inequality \cite{CHSH69}.

The bounds in inequality (\ref{Mermins}) can be easily derived as follows:
In any local-realistic theory, for any individual system, observables
$A_j$ and $B_j$ have predefined values $a_j$ and $b_j$, respectively.
Each of these values is constrained to lie between $-s$ and $s$.
Since $M_n^{(s)}$ is linear in each local observable (fixing the value of the
other $2 n-1$ local observables), $M_n^{(s)}$ will take its extremal values when
local observables take their extremal values, $-s$ or $s$. The various
combinations of $a_j=\pm s$ and $b_j=\pm s$ always give
$\pm 2^{n - 1} s^n$, Q.E.D.


\section{Violations of the
generalized Mermin-Klyshko inequality}


For a $n$ spin-$s$ particle system in
a quantum pure state
$\left| \psi \right\rangle$, the quantum correlation of $A_1$,
\ldots, $A_n$ is defined as
$\left\langle \psi \right| \hat A_1^{(s)} \otimes \cdots \otimes
\hat A_n^{(s)} \left| \psi \right\rangle$,
where $\hat A_1^{(s)}$, \ldots, $\hat A_n^{(s)}$ are
the self-adjoint operators that represent the local observables $A_1^{(s)}$,
\ldots, $A_n^{(s)}$.

Let us consider the following local operators on particle $j$:
\begin{eqnarray}
\hat A_j^{(s)} & = & \left( {\matrix{
s&{}&{}&{}&{}\cr
{}&{s-1}&{}&{}\cr
{}&{}&\cdots&{}&{}\cr
{}&{}&{}&{-s+1}&{}\cr
{}&{}&{}&{}&{-s}\cr
}} \right),
\label{Aop} \\
\hat B_j^{(s)} & = & \left( {\matrix{
{}&{}&{}&{}&s\cr
{}&{}&{}&{s-1}&{}\cr
{}&{}&\cdots&{}&{}\cr
{}&{s-1}&{}&{}&{}\cr
s&{}&{}&{}&{}\cr
}} \right).
\label{Bop}
\end{eqnarray}
$\hat A_j^{(s)}$ and $\hat B_j^{(s)}$ are diagonal $(2 s+1) \times (2 s+1)$ matrices,
with nondegenerated eigenvalues $-s$, $-s+1$, $\ldots$, $s-1$, $s$.

In addition, let us recursively define the following operator on the composite
system consisting on $n \ge 2$ spin-$s$
particles:
\be
\hat M_n^{(s)}=\hat M_{n-1}^{(s)} \otimes
\left(\hat A_n^{(s)}+\hat B_n^{(s)}\right) +
\hat K_{n-1}^{(s)} \otimes
\left(\hat A_n^{(s)}-\hat B_n^{(s)}\right),
\ee
letting $\hat M_1^{(s)}=\hat A_1^{(s)}$, and $\hat K_n^{(s)}$ being
the same as $\hat M_n^{(s)}$ but exchanging
the $\hat A$'s for $\hat B$'s.

As can be easily checked, $\hat M_n^{(s)}$ is a linear combination
of $2^{2 f\left({n / 2}\right)}$
operators of the type $\hat A_1^{(s)} \otimes \cdots \otimes \hat A_n^{(s)}$
(all of them commuting if $n$ is odd, but not if it is even).
The greatest eigenvalue of $\hat M_n^{(s)}$ is $2^{3 (n-1) / 2} s^n$, which is nondegenerated.
Let us consider the corresponding eigenstate
$\left|\,{\mu}_n^{(s)} \right\rangle$, characterized by
the equation
\begin{equation}
\hat M_n^{(s)} \left|\,{\mu}_n^{(s)} \right\rangle =
2^{3 (n-1) / 2} s^n \left|\,{\mu}_n^{(s)} \right\rangle.
\label{eigenvalueeq}
\end{equation}

For $n=2$,
$\left|\,\mu_n^{(s)} \right\rangle$ is a maximally entangled state
of two spin-$s$ particles. For $n \ge 3$, $\left|\,\mu_n^{(s)} \right\rangle$
is a generalized GHZ state,
as defined in Ref.~\cite{Cabello01}, and allows us to develop an
EPR-like argument for observables $A_j$ and $B_j$. For $n$ odd (even),
$\left|\,\mu_n^{(s)} \right\rangle$ and all the operators (a subset of
mutually commuting operators)
of the type $\hat A_1^{(s)} \otimes \cdots \otimes \hat A_n^{(s)}$
included in $\hat M_n^{(s)}$
allow us to develop a GHZ-like proof without inequalities of Bell's theorem
\cite{com1}
(see \cite{Cabello01} for the details).

In this paper, however, we are interested in
violations of inequality (\ref{Mermins}).
For that purpose, let us take a look at
the prediction of quantum mechanics for
the state $\left|\,{\mu}_n^{(s)} \right\rangle$
for the
combination of correlations appearing in inequality (\ref{Mermins}).
Observable $M_n^{(s)}$ is represented in
quantum mechanics by the self-adjoint operator $\hat M_n^{(s)}$.
Therefore, as can be immediately seen in Eq.~(\ref{eigenvalueeq}),
according to quantum mechanics the expected value for $M_n^{(s)}$
in the state $\left|\,{\mu}_n^{(s)} \right\rangle$ is given by
\be
\left\langle \mu_n^{(s)} \right| \hat M_n^{(s)}
\left| \,\mu_n^{(s)} \right\rangle
= 2^{3 (n-1) / 2} s^n.
\label{qprediction}
\ee
This value violates inequality (\ref{Mermins}).
Indeed, it can be proved that
this is the maximum allowed violation of
inequality (\ref{Mermininequalitys}).
The proof is simple for $n$ odd. Then, $\hat M_n^{(s)}$
is a linear combination with coefficients $\pm 2^{(n-1)/2}$
of $2^{n-1}$ operators of the type
$\hat A_1^{(s)} \otimes \cdots \otimes \hat A_n^{(s)}$, and
each of these correlations is bound by $\pm s^n$.
Therefore, for $n$ odd,
the maximum value that $M_n^{(s)}$ can reach is, by definition,
$2^{3 (n-1) / 2} s^n$, Q.E.D.

If $n$ is even the proof is more difficult
(for $n=2$ and $s=1$, or for $s={1 \over 2}$ and choosing units in which $2 \hbar =1$,
proofs can be found in Refs.~\cite{Cirelson80,Landau87}).


\section{Conclusions}


The ratio between the quantum
correlation given by Eq.~(\ref{qprediction}) and the maximal classical one,
which appears in Eq.~(\ref{Mermininequalitys}), is
\be
{ \left\langle \mu_n^{(s)} \right| \hat M_n^{(s)}
\left| \,\mu_n^{(s)} \right\rangle
\over \mbox{max\,} M_n^{(s)}} =
2^{(n - 1) / 2}\;\;\forall s.
\label{Merminratio}
\ee
That is, for a fixed $n \ge 2$ the contradiction between quantum mechanics
and local realism is {\em constant} as the spin $s$ increases.
For $n=2$ the same conclusion was reached by Gisin and Peres in Ref.~\cite{GP92}.
Therefore, our analysis is in agreement with Gisin and Peres's
and generalizes it to systems of $n \ge 2$ particles.

On the other hand, ratio (\ref{Merminratio}) shows that for a fixed $s$, the
correlations found by $n$ distant observers violate the classical bound by
a factor that increases
{\em exponentially} with the number $n$ of particles.
For $s = {1 \over 2}$ the same conclusion was reached by Mermin in Ref.~\cite{Mermin90b}.
Thus our analysis generalizes Mermin's to
systems of spin $s \ge {1 \over 2}$.

Therefore, the approach presented in this paper
unifies and generalizes some previous results,
in particular, those in Refs.~\cite{GP92,Mermin90b,BK93}, and
unifies the conclusions reached in Refs.~\cite{Peres92,GP92,Mermin90b}:
Neither a large spin nor
a large number of particles nor a large number of large spin particles
guarantee classical behavior.

In addition, this approach allows us to translate the proofs
without inequalities of Bell's theorem
for multiparty multilevel GHZ states introduced in Ref.~\cite{Cabello01}
into Bell's inequalities that can be tested in real experiments.


\begin{acknowledgments}
The author thanks J. L. Cereceda, C. Serra, and M. M. Wolf for comments, and
the Spanish Ministerio de Ciencia y Tecnolog\'{\i}a
Grant No. BFM2001-3943 and
the Junta de Andaluc\'{\i}a
Grant No. FQM-239 for support.
\end{acknowledgments}



\begin{thebibliography}{99}
\bibitem{com1} The operators of the type
$\hat A_1^{(s)} \otimes \cdots \otimes \hat A_n^{(s)}$
appearing in
Mermin's original inequality for $n$ spin-${1 \over 2}$ particles
in Ref.~\cite{Mermin90b} are mutually compatible
both for $n$ odd and even, and can be used to develop
a GHZ-like proof of Bell's theorem without inequalities in
both cases.
\end{thebibliography}
\end{document}